\documentstyle[12pt,a4wide,pstcol,epsf]{article}

\newcommand{\lsim}{\mbox{\raisebox{-.9ex}{~$\stackrel{\mbox{$<$}}{\sim}$~}}}
\newcommand{\gsim}{\mbox{\raisebox{-.9ex}{~$\stackrel{\mbox{$>$}}{\sim}$~}}}

\def\thebiblio#1{
\begin{center}\bf \large References
\end{center}
\list
{[\arabic{enumi}]}{\settowidth\labelwidth{#1.}\leftmargin\labelwidth
 \advance\leftmargin\labelsep
 \usecounter{enumi}}
 \def\newblock{\hskip .11em plus .33em minus -.07em}
 \sloppy
 \sfcode`\.=1000\relax}

\newcommand{\caln}{{\cal N}}

\newcommand{\calp}{{\cal P}}
\newcommand{\cala}{{\cal A}}

\begin{document}


\begin{center}

{\Large\bf Statistical Anisotropy and the\\ Vector Curvaton Paradigm}

\bigskip

{\large Konstantinos~Dimopoulos}
\footnote{k.dimopoulos1@lancaster.ac.uk}

\medskip

{\it Physics Department, Lancaster University, Lancaster LA1 4YB, UK}

\date{\today}

%
\begin{abstract}
The vector curvaton paradigm is reviewed. The mechanism allows a massive vector
boson field to contribute to or even generate the curvature perturbation in the
Universe. Contribution of vector bosons is likely to generate statistical 
anisotropy in the spectrum and bispectrum of the curvature perturbation, which 
will soon be probed observationally. Two specific models for the generation of
superhorizon spectra for the components of an Abelian vector field are 
analysed. Emphasis is put on the observational signatures of the models when 
the vector fields play the role of vector curvatons. If future observations 
support the vector curvaton mechanism this will open a window into the gauge
field content of theories beyond the standard model.
\end{abstract}



\end{center}


\section{Introduction}

Cosmic inflation is arguably the most compelling way to overcome, or at least
ameliorate, the so-called horizon and flatness problems of the hot big bang 
cosmology. However, these problems are successfully addressed by all models
of inflation, provided that the inflationary expansion lasts long enough. 
Therefore, discrimination between inflation models is based on another 
important consequence of inflationary expansion, namely the generation of
the curvature perturbation $\zeta$ in the Universe, which is responsible for
the formation of structures such as galaxies and galactic clusters and which
is reflected onto the Cosmic Microwave Background (CMB) radiation through the 
Sachs-Wolfe effect. 

The latest CMB observations appear to confirm the ``vanilla'' predictions of
inflation with respect to $\zeta$: scale-invariance, Gaussianity and 
statistical homogeneity and isotropy. However, a period of accelerated 
expansion
of space (this is the definition of cosmic inflation) is not enough to 
guarantee scale-invariance for the curvature perturbation. Indeed, inflation
is required to be of the quasi-de Sitter type, where the density of the 
Universe remains roughly constant. Also, the CMB observations seem to suggest
that exact scale-invariance is not favoured (although it is not ruled out) 
with the spectral index of $\zeta$, satisfying \mbox{$n_s-1=-0.037\pm0.014$}
(at 1$\sigma$) \cite{wmap7} when $\Lambda$CDM cosmology is assumed.%
\footnote{Exact scale invariance corresponds to \mbox{$n_s=1$}.} Most likely,
this reveals some dynamics for inflation (the density is not exactly constant),
which is indeed expected by model-builders. Similarly, a high degree of 
Gaussianity in the curvature perturbation is expected
since this reflects the randomness of quantum fluctuations, on which
the particle production process is feeding, in order to generate the 
superhorizon spectrum of the fields which eventually give rise to $\zeta$.
However, the Gaussianity of $\zeta$ crucially depends on the linearity of the
process which translates these field perturbations to $\zeta$. This is 
quantified through the so-called non-linearity parameter $f_{\rm NL}$. The 
latest CMB observations provide a hint of non-zero non-Gaussianity since, in 
the squeezed configuration, \mbox{$f_{\rm NL}=32\pm21$} (at 1$\sigma$) 
\cite{wmap7}, which again
appears to deviate from the ``vanilla'' prediction.\footnote{Note that $\zeta$
is indeed predominantly Gaussian because the bispectrum $B_\zeta$ is related
to the power spectrum $P_\zeta$ roughly as 
\mbox{$B_\zeta\sim f_{\rm NL}P_\zeta^2$} (see Eq.~(\ref{fNLdef})), with the 
observations suggesting \mbox{$\calp_\zeta=4.8\times 10^{-5}$} for the spectrum
\cite{wmap7}.}

In the same spirit, observations suggest that there may be deviations from
statistical homogeneity, since there seem to be a difference in the power of 
$\zeta$ as large as 10\% between hemispheres in the CMB \cite{stinhom}. 
Finally, statistical isotropy in $\zeta$ is also questioned by observations.
Indeed, there is tantalising evidence of a preferred direction on
the microwave sky. This is the so-called ``Axis of Evil'' observation 
\cite{AoE}, which amounts to an alignment of the quadrupole and octupole
moments in the CMB, which is statistically extremely unlikely \cite{anistat}
and has been shown to persist beyond foreground removal \cite{forg}.

Thus, we see that the precision of the cosmological observations is such that
begins to enable us to explore beyond the ``vanilla'' predictions of inflation
and use the observed deviations from them to discriminate between classes of 
models and paradigms. There is already a huge literature on deviations from
exact scale-invariance and Gaussianity for the curvature perturbation.
In this paper we discuss possible deviations from statistical isotropy and
we present a compelling paradigm for their generation; namely the Vector 
Curvaton Paradigm.

Throughout the paper we consider a metric with negative signature and use
natural units where \mbox{$c=\hbar=k_B=1$} and Newton's gravitational constant 
is \mbox{$8\pi G=m_P^{-2}$}, with \mbox{$m_P=2.4\times 10^{18}\,$GeV} being the
reduced Planck mass.

\section{Statistical anisotropy in the curvature perturbation}

Statistical anisotropy amounts to direction dependent patterns in the CMB. It 
can be quantified as follows. The spectrum $\calp_\zeta$ of the curvature 
perturbation is defined through the two-point correlator as
\begin{equation}
\langle\zeta(\mbox{\boldmath $k$})\zeta(\mbox{\boldmath $k$}')\rangle=
(2\pi)^3\delta(\mbox{\boldmath $k$}+\mbox{\boldmath $k$}')\frac{2\pi^2}{k^3}
\calp_\zeta(\mbox{\boldmath $k$}),
\end{equation}
where \mbox{$k=|\mbox{\boldmath $k$}|$} and 
$$\zeta(\mbox{\boldmath $k$})\equiv\int\zeta(\mbox{\boldmath $x$})
e^{-i\mbox{\scriptsize\boldmath $k\cdot x$}}d^3x\,.$$
The reality condition 
\mbox{$\zeta^*(\mbox{\boldmath $k$})=\zeta(-\mbox{\boldmath $k$})$}
demands that 
\mbox{$\calp_\zeta(\mbox{\boldmath $k$})=\calp_\zeta(-\mbox{\boldmath $k$})$}. 
Now, the dependence of the power spectrum on the direction of the momentum 
vector can be parametrised as~\cite{ACW}
\begin{equation}
\calp_\zeta(\mbox{\boldmath $k$})=\calp_\zeta^{\rm iso}(k)
[1+g(\mbox{\boldmath $\hat d\cdot\hat k$})^2+\cdots],
\label{gdef}
\end{equation}
where {\boldmath $\hat d$} is the unit vector along the preferred direction,
\mbox{\boldmath $\hat k\equiv k$}$/k$ and the ellipsis denotes higher than 
quadratic order terms, which are negligible if $g<1$. \mbox{$g=g(k)$} is 
sometimes called the anisotropy parameter. A similar 
parametrisation can be assigned to higher order correlators, i.e. the 
bispectrum, trispectrum etc. (for the bispectrum see Sec.~\ref{bi}).

What are observations saying about $g$? In Ref.~\cite{GE} the latest CMB data 
was analysed without the prior of statistical isotropy,
in order to obtain bounds or even detect statistical anisotropy. Indeed, it was
found that \mbox{$g=0.29\pm 0.03$} at the level of 9$\sigma$!. However, the 
preferred direction was too close to the ecliptic plane so the authors 
suspected some unknown systematic.\footnote{It is important to note that this
finding was also confirmed by the WMAP team in Ref.~\cite{wmap7anomalies}, who 
suspect that it is due to beam assymmetries but have not fully expalined it 
yet.} Hence, this number can be considered only as an upper bound 
\mbox{$g\lsim 0.3$}. The observations of the Planck satellite 
will strengthen this bound by at least an order of magnitude and reduce it to
\mbox{$g\lsim 0.02$} if statistical anisotropy in the spectrum of the curvature
perturbation is indeed not observed \cite{planck}.

\section{Vector fields and the curvature perturbation}

Statistical anisotropy in the curvature perturbation cannot be generated if
one considers the effects of scalar fields only, because the latter cannot
introduce a preferred direction. In this paper we will study how vector boson 
fields can directly influence the curvature perturbation and generate 
statistical anisotropy.\footnote{Indirectly, statistical anisotropy in $\zeta$
can also be generated by considering a mild anisotropisation of the 
inflationary expansion, due to the presence of a vector boson field condensate.
In this case, it is the perturbations of the inflaton scalar field which are
rendered statistically anisotropic \cite{anisinf}.} 
We will investigate the contribution of vector fields to $\zeta$
through the so-called $\delta N$ formalism \cite{dN}.

According the $\delta N$ formalism the curvature perturbation is the
difference of the logarithmic expansion between uniform density and 
spatially flat slices of spacetime: \mbox{$\zeta=\delta(\ln a)\equiv\delta N$},
where $a$ is the scale factor of the Universe and $N$ corresponds to the 
elapsing e-folds of expansion. We will assume that $N$ is
influenced by both scalar and vector boson fields. For simplicity, we consider
one of each of such fields\footnote{For the multifield case see 
Ref.~\cite{Y1loop}.}, i.e. \mbox{$N=N(\phi,\mbox{\boldmath $A$})$}. 
Then, the curvature perturbation can be written as
\begin{equation}
\zeta=N_\phi\delta\phi+N_A^i\delta A_i+\frac12 N_{\phi\phi}(\delta\phi)^2+
\frac12 N_{\phi A}^i\delta\phi\delta A_i
+\frac12 N_{AA}^{ij}\delta A_i\delta A_j+\cdots\,,
\label{dN}
\end{equation}
where \mbox{$N_\phi\equiv\frac{\partial N}{\partial\phi}$},
\mbox{$N_A^i\equiv\frac{\partial N}{\partial A_i}$},
\mbox{$N_{\phi\phi}\equiv\frac{\partial^2 N}{\partial\phi^2}$}
\mbox{$N_{\phi A}^i\equiv\frac{\partial^2 N}{\partial\phi\partial A_i}$}
and \mbox{$N_{AA}^{ij}\equiv\frac{\partial^2 N}{\partial A_i\partial A_j}$}, 
with \mbox{$i=1,2,3$} labelling spatial 
components and Einstein summation over repeated indexes is assumed. Now, 
since the vector field has three degrees of freedom, at a flat slice of
spacetime foliation, we define
\begin{equation}
\delta\mbox{\boldmath $A$}(\mbox{\boldmath $k$}, t)=
\sum_\lambda\mbox{\boldmath $\hat e$}_\lambda(\mbox{\boldmath $\hat k$})
\delta A_\lambda(\mbox{\boldmath $k$}, t)\,,
\end{equation}
where \mbox{$\lambda=L,R,\|$} denotes the three polarisations and 
the polarisation vectors can be defined as
\begin{equation}
\hat e_L\equiv\frac{1}{\sqrt 2}(1,i,0),\quad
\hat e_R\equiv\frac{1}{\sqrt 2}(1,-i,0)\quad{\rm and}\quad
\hat e_\|\equiv(0,0,1)\,,
\end{equation}
where `L', `R' denote the left and right transverse polarisations respectively 
and `$\|$' denotes the longitudinal polarisation (if physical).
Then, assuming approximately isotropic expansion\footnote{%
This is in contrast to Ref.~\cite{anisinf} where it is the anisotropy in the 
expansion which sources statistical anisotropy in $\zeta$.}, the power-spectrum
for each polarisation of the vector field perturbations is
\begin{equation}
\langle\delta A_\lambda(\mbox{\boldmath $k$})
\delta A_\lambda(\mbox{\boldmath $k$}')\rangle=
(2\pi)^3\delta(\mbox{\boldmath $k$}+\mbox{\boldmath $k$}')\frac{2\pi^2}{k^3}
\calp_\lambda(k)\,.
\end{equation}

\subsection{The spectrum}

In Ref.~\cite{stanis}, it was shown that the correlators of the perturbations
of the vector field can be written as
\begin{equation}
\langle\delta A_i(\mbox{\boldmath $k$})
\delta A_j(\mbox{\boldmath $k$}')\rangle=
(2\pi)^3\delta(\mbox{\boldmath $k$}+\mbox{\boldmath $k$}')\frac{2\pi^2}{k^3}
\left[T_{ij}^{+}(\mbox{\boldmath $\hat k$})\calp_++
iT_{ij}^{-}(\mbox{\boldmath $\hat k$})\calp_-+
T_{ij}^\|(\mbox{\boldmath $\hat k$})\calp_\|\right],
\end{equation}
where
\begin{equation}
T_{ij}^{+}(\mbox{\boldmath $\hat k$})\equiv\delta_{ij}-\hat k_i\hat k_j\;,\quad
T_{ij}^{-}(\mbox{\boldmath $\hat k$})\equiv\varepsilon_{ijk}\hat k_k
\quad{\rm and}\quad
T_{ij}^\|(\mbox{\boldmath $\hat k$})\equiv\hat k_i\hat k_j\;
\end{equation}
with $\delta_{ij}$ being the Kronecker's delta,
and we have defined for the transverse spectra
\begin{equation}
\calp_\pm\equiv\frac12(\calp_R\pm\calp_L)\,,
\end{equation}
denoting the parity even and odd polarisations. For a parity conserving theory
we have \mbox{$\calp_-=0$}.
Using the above, we obtain the power spectrum of the curvature perturbation as
\begin{eqnarray}
\calp_\zeta(\mbox{\boldmath $k$}) & = & N_\phi^2\calp_\phi(k)+
N_A^iN_A^j\left[
T_{ij}^{+}(\mbox{\boldmath $\hat k$})\calp_+(k)+
T_{ij}^\|(\mbox{\boldmath $\hat k$})\calp_\|(k)\right]\;=\nonumber\\
 & = &  N_\phi^2\calp_\phi+N_A^2\left[
\calp_++(\calp_\|-\calp_+)(\mbox{\boldmath $\hat N_A\cdot\hat k$})^2\right],
\end{eqnarray}
where \mbox{$N_A\equiv|\mbox{\boldmath $N_A$}|=\sqrt{N_A^iN_A^i}$}, 
\mbox{$\mbox{\boldmath $\hat N_A$}\equiv\mbox{\boldmath $N_A$}/N_A$}. 

From the above we see that the isotropic part of the spectrum is
\begin{equation}
\calp_\zeta^{\rm iso}(k)=N_\phi^2\calp_\phi(k)+N_A^2\calp_+(k)
\label{Piso}
\end{equation}
and the preferred direction is given by 
\mbox{$\mbox{\boldmath $\hat d$}=\mbox{\boldmath $\hat N_A$}$} 
(cf. Eq.~(\ref{gdef})). The anisotropy parameter is
\begin{equation}
g=\frac{N_A^2(\calp_\|-\calp_+)}{N_\phi^2\calp_\phi+N_A^2\calp_+}
=\beta\frac{\calp_\|-\calp_+}{\calp_\phi+\beta\calp_+}\,,
\label{g}
\end{equation}
where we have defined
\begin{equation}
\beta\equiv\frac{N_A^2}{N_\phi^2}\,,
\label{beta}
\end{equation}
which quantifies the relative contribution of the vector over the scalar field
to the modulation of $N$. Notice that particle production becomes isotropic
(\mbox{$g=0$}) if \mbox{$\calp_+=\calp_\|$}.

\subsection{The bispectrum}\label{bi}

Statistical anisotropy is also possible to be manifest in higher order 
correlators of the curvature perturbation. In this paper we discuss only the 
bispectrum (for the trispectrum see Ref.~\cite{Ytri}).

The bispectrum of the curvature perturbation is defined as
\begin{equation}
\langle\zeta(\mbox{\boldmath $k$})
\zeta(\mbox{\boldmath $k$}')
\zeta(\mbox{\boldmath $k$}'')\rangle=(2\pi)^3
\delta(\mbox{\boldmath $k$}+\mbox{\boldmath $k$}'+\mbox{\boldmath $k$}'')
B_\zeta(\mbox{\boldmath $k$},\mbox{\boldmath $k$}',\mbox{\boldmath $k$}'')\,.
\label{zB}
\end{equation}
The bispectrum $B_\zeta$ is a measure of the non-Gaussianity of the curvature
perturbation since, for Gaussian $\zeta$, $B_\zeta$ is exactly zero. 

The curvature perturbation is generated due to the quantum fluctuations of 
suitable fields which are stretched to become classical perturbations during 
inflation. Since quantum fluctuations are Gaussian (which reflects their 
randomness) sizable non-Gaussianity in $\zeta$ is generated only
if the process through which the perturbations of the relevant fields affect
the expansion of the Universe and imprint their contribution to the curvature 
perturbation. If this process is significantly non-linear deviations from 
Gaussianity will be generated. This is why the bispectrum is quantified
by the so-called non-linearity parameter $f_{\rm NL}$, which can be defined as 
follows
\begin{equation}
B_\zeta(\mbox{\boldmath $k_1$},\mbox{\boldmath $k_2$},\mbox{\boldmath $k_3$})=
\frac65 f_{\rm NL}
\left[P_\zeta(k_1)P_\zeta(k_2)+
P_\zeta(k_2)P_\zeta(k_3)+
P_\zeta(k_3)P_\zeta(k_1)
\right],
\label{fNLdef}
\end{equation}
where \mbox{$4\pi k^3 P_\zeta\equiv(2\pi)^3\calp_\zeta$}. The value of 
$f_{\rm NL}$ depends on the configuration of the three momentum vectors 
which are used to define the bispectrum. The most popular configurations
are the ``equilateral'', for which \mbox{$k_1=k_2=k_3$}, and the 
``squeezed'', for which \mbox{$k_1=k_2\gg k_3$}.

How does the contribution of a vector field affect the bispectrum of the 
curvature perturbation? In Ref.~\cite{fnlanis} it was shown that
\begin{equation}
B_\zeta=B_\phi+B_{\phi A}+B_A\;,
\end{equation}
where
\begin{equation}
B_\phi=N_\phi^2 N_{\phi\phi}\left[
\frac{4\pi^4}{k_1^3k_2^3}\calp_\phi(k_1)\calp_\phi(k_2)+
\frac{4\pi^4}{k_2^3k_3^3}\calp_\phi(k_2)\calp_\phi(k_3)+
\frac{4\pi^4}{k_3^3k_1^3}\calp_\phi(k_3)\calp_\phi(k_1)\right],
\end{equation}
\begin{equation}
B_{\phi A}=-\frac12 N_\phi N_{\phi A}^i\left[
\frac{4\pi^4}{k_1^3k_2^3}\calp_\phi(k_1){\cal M}_i(\mbox{\boldmath $k_2$})+
{\rm 5\;cyclic\;permutations}\,\right]
\end{equation}
and
\begin{equation}\,\!
B_A\!=\!\frac{4\pi^4}{k_1^3k_2^3}{\cal M}_i(\mbox{\boldmath $k_1$})N_{AA}^{ij}
{\cal M}_j(\mbox{\boldmath $k_2$})\!+\!
\frac{4\pi^4}{k_2^3k_3^3}{\cal M}_i(\mbox{\boldmath $k_2$})N_{AA}^{ij}
{\cal M}_j(\mbox{\boldmath $k_3$})\!+\!
\frac{4\pi^4}{k_3^3k_1^3}{\cal M}_i(\mbox{\boldmath $k_3$})N_{AA}^{ij}
{\cal M}_j(\mbox{\boldmath $k_1$}),\hspace{-1cm}
\end{equation}
where
\begin{equation}
\mbox{\boldmath ${\cal M}$}(\mbox{\boldmath $k$})\equiv\calp_+(k)N_A\left[
\mbox{\boldmath $\hat N_A$}+p(k)\mbox{\boldmath $\hat k$}
(\mbox{\boldmath $\hat k\cdot\hat N_A$})+
iq(k)\mbox{\boldmath $\hat k\times\hat N_A$}\right]
\label{calM}
\end{equation}
and we have defined
\begin{equation}
p\equiv\frac{\calp_\|-\calp_+}{\calp_+}\quad{\rm and}\quad
q\equiv\frac{\calp_-}{\calp_+}\,.
\label{pq}
\end{equation}
Using the above, we obtain $f_{\rm NL}$ in the equilateral and squeezed 
configurations respectively as follows
\begin{equation}
\frac65 f_{\rm NL}^{\rm eql}=\frac{{\cal B}_\zeta^{\rm eql}
(\mbox{\boldmath $k_1$},\mbox{\boldmath $k_2$},\mbox{\boldmath $k_3$})}{3
[\calp_\zeta^{\rm iso}(k)]^2}
\label{fnleql}
\end{equation}
and
\begin{equation}
\frac65 f_{\rm NL}^{\rm sqz}=\frac{{\cal B}_\zeta^{\rm sqz}
(\mbox{\boldmath $k_1$},\mbox{\boldmath $k_2$},\mbox{\boldmath $k_3$})}{2
\calp_\zeta^{\rm iso}(k_1)\calp_\zeta^{\rm iso}(k_3)}\,,
\label{fnlsqz}
\end{equation}
where 
\mbox{${\cal B}_\zeta^{\rm eql}\equiv
\left(\frac{k^3}{2\pi^2}\right)^2B_\zeta^{\rm eql}$}
and
\mbox{${\cal B}_\zeta^{\rm sqz}\equiv
\frac{k_1^3k_3^3}{4\pi^4}B_\zeta^{\rm sqz}$}, with \mbox{$k_1=k_2\gg k_3$} in
the squeezed configuration and 
\mbox{$k\equiv k_1=k_2=k_3$} in the equilateral configuration.


\section{The Vector Curvaton Paradigm}

For a vector field to directly affect the curvature perturbation in the 
Universe we need two ingredients. First, we need a mechanism to break the 
conformal invariance of the vector field and generate a superhorizon spectrum 
of vector field perturbations $\delta A_\mu$. 
Second, we need a 
mechanism that will allow these perturbations to affect (or even generate) the 
curvature perturbation $\zeta$. This can be done only if the vector field 
and/or its perturbations, in some way affect the Universe evolution.

In this section we focus on the second ingredient, i.e. on a mechanism for the
generation of a contribution from the vector field perturbations to the 
curvature perturbation of the Universe; namely the Vector Curvaton mechanism. 
Thus, we assume that some given mechanism has produced the necessary 
superhorizon spectrum of vector field perturbations (as is discussed in 
Sec.~\ref{pp}) during inflation, which for the moment we take for granted. 

A single vector field cannot play the role of the inflaton. The reason 
is straightforward. Inflation homogenises a vector field and a homogeneous
vector field picks up a preferred direction in space.\footnote{Unless one
tunes the spatial components of the vector field to zero by design
\cite{spatial}.} Thus, if a homogeneous
vector field dominated the Universe during inflation it would lead to excessive
anisotropic stress, which would produce too much of a large-scale anisotropy
and, therefore, will be in conflict with CMB observations. A huge number 
$\caln$ of vector fields, randomly oriented, could avoid this problem 
\cite{VI}. Indeed, if this is the case then the statistical anisotropy produced
is \mbox{$g\propto 1/\sqrt\caln$}, which means that hundreds of 
vector fields are needed to satisfy the observational bounds. This not only 
implies the use of giant gauge groups but also requires the tuning of the 
initial conditions so that they are the same for all the fields. Another option
is to consider a ``triad'' of orthogonally oriented vector fields (again with 
the same initial conditions) so that the excessive anisotropic stress is 
eliminated \cite{triad}. For the above reasons we will not consider vector 
fields as inflatons.

If the vector field is not the inflaton, it needs to affect the Universe 
expansion in some other way, either at the end or after the end of inflation.
There are a multitude of mechanisms which may allow a vector field to do that,
mirroring the corresponding scalar field models. Prominent examples include
the curvaton \cite{curv}, the inhomogeneous end of inflation \cite{endinf} (see
also Ref.~\cite{endinf+}) and the modulated reheating \cite{modreh} mechanisms.
Historically, statistical anisotropy by vector field perturbations was first 
studied in the context of the inhomogeneous end of inflation mechanism 
\cite{yokosoda},%
\footnote{For modulated reheating with vector fields see Ref.~\cite{sugravec}.}
using a particular model of hybrid inflation.%
\footnote{For non-Gaussianity in this model see also \cite{yokosoda+}.}
Here, however, we concentrate on the curvaton mechanism, which has the 
considerable advantage that it does not rely on an interaction of any kind 
between the vector field and the inflaton sectors. This avoids overcomplicating
the model but, more importantly, allows the vector curvaton sector to be 
completely independent of the physics of inflation. As such, not only can it 
apply in many given inflation scenarios, but it can correspond to physics at
a much lower energy scale than inflation; even TeV physics. As is 
the case of the scalar curvaton, the vector curvaton is not a particular model 
but it can correspond to a multitude of 
realisations, hence we refer to the mechanism as a paradigm rather than a 
model. The vector curvaton mechanism was first introduced in the pioneering 
work in Ref.~\cite{vecurv}, which was the first article to consider the 
possibility that a vector boson field can contribute to the curvature 
perturbation in the Universe.

The idea of the curvaton assumes the existence of a spectator field during 
inflation, which has nothing to do with inflationary dynamics but is light 
enough so that it manages to obtain a superhorizon spectrum of perturbations. 
After the end of inflation (possibly long afterwards), the curvaton becomes
heavy and begins undergoing oscillations which allow it to come to dominate (or
nearly dominate) the Universe before its decay. Owing to its perturbations, the
density of the curvaton is perturbed throughout space so that its (near) 
domination occurs at different times at different locations. Thus, its effect
on the evolution of the Universe is location dependent, which is the reason why
it can affect (or even generate) the curvature perturbation in the Universe. 
Note that, for a vector field to do this, it must avoid generating an 
excessive anisotropic stress at domination.

\subsection{The setup}

Consider a massive Abelian vector boson field, with Lagrangian density
\begin{equation}
{\cal L}=-\frac14 F_{\mu\nu}F^{\mu\nu}+\frac12 m^2W_\mu W^\mu,
\label{L0}
\end{equation}
where $W_\mu$ is the vector field, $m$ is its mass and
\mbox{$F_{\mu\nu}=\partial_\mu W_\nu-\partial_\nu W_\mu$} is the field
strength tensor. Inflation homogenises the vector field so that 
\mbox{$W_\mu=W_\mu(t)$}. If \mbox{$m\neq 0$} it is easy to show that the 
temporal component of the homogeneous vector field is zero, i.e. 
\mbox{$W_t=0$}.
If \mbox{$m=0$} then the field is gauge invariant and we can set \mbox{$W_t=0$}
by virtue of a gauge choice. However, in this case the value of the spatial 
vector field {\boldmath $W$} is not well defined because gauge invariance 
allows us to change it as 
\mbox{{\boldmath $W$}$\rightarrow${\boldmath $W$}$+${\boldmath $C$}}, where 
{\boldmath $C$} is a constant vector of arbitrary magnitude. Thus, we will
concentrate on the case \mbox{$m\neq 0$} from now on, where gauge invariance 
is broken and the homogeneous ``zero-mode'' is well defined.

The energy-momentum tensor for the vector field is
\begin{equation}
T_{\mu\nu}=\frac14 g_{\mu\nu}F_{\rho\sigma}F^{\rho\sigma}-F_{\mu\rho}F_\nu^\rho
+m^2\left(W_\mu W_\nu-\frac12 g_{\mu\nu}W_\rho W^\rho\right),
\end{equation}
where $g_{\mu\nu}$ is the metric tensor (negative signature is assumed).
The above can be written as \cite{vecurv}
\begin{equation}
T_\mu^\nu={\rm diag}(\rho_A, -p_\perp, -p_\perp, +p_\perp)\,,
\label{Tdiag}
\end{equation}
where
\begin{equation}
\rho_A\equiv\rho_{\rm kin}+V_A\quad{\rm and}\quad
p_\perp\equiv\rho_{\rm kin}-V_A
\label{rkinp}
\end{equation}
with 
\begin{equation}
\rho_{\rm kin}\equiv-\frac14 F_{\mu\nu}F^{\mu\nu}\quad{\rm and}\quad
V_A\equiv-\frac12 m^2W_\mu W^\mu,
\label{rkinVA}
\end{equation}
where `kin' denotes the kinetic density $\rho_{\rm kin}$.
Notice that the energy-momentum tensor is similar to the one of a perfect fluid
with the crucial difference that the pressure along the longitudinal direction
is of opposite sign to the pressure along the transverse directions. This means
that, if this pressure were not zero and the vector field dominated the 
Universe, it would give rise to significant anisotropic stress, which is the 
reason why a single vector field cannot play the role of the inflaton.

Using Eq.~(\ref{L0}) one can obtain the equation of motion for the homogeneous
vector field, which reads
\begin{equation}
\mbox{\boldmath $\ddot W$}+H\mbox{\boldmath $\dot W$}+m^2\mbox{\boldmath $W$}
=0\,,
\label{eomW}
\end{equation}
where the dot denotes the derivative with respect to the cosmic time $t$ and 
\mbox{$H\equiv\dot a/a$} is the Hubble parameter, i.e. the rate of the Universe
expansion. At this
point we need to stress that \mbox{\boldmath $W$} is the {\em comoving} and not
the physical vector field. Indeed, the mass term in Eq.~(\ref{L0}) can be 
written as
\begin{equation}
\delta{\cal L}_m\equiv \frac12m^2W_\mu W^\mu=\frac12 m^2(W_t^2-a^{-2}W_iW_i)=
-\frac12 m^2|\mbox{\boldmath $W$}/a|^2,
\end{equation}
where we used that \mbox{$W_t=0$} and a spatially flat FRW metric
\mbox{$ds^2=dt^2-a^2dx^idx^i$}. From the 
above it can deduced that the {\em physical} vector field has spatial 
components
\begin{equation}
\mbox{\boldmath $A$}\equiv\mbox{\boldmath $W$}/a\,.
\label{phys}
\end{equation}

In terms of the physical vector field, Eq.~(\ref{eomW}) is written as
\begin{equation}
\ddot A+3H\dot A+(\dot H+2H^2+m^2)A=0\,,
\label{eomhom}
\end{equation}
where \mbox{$A\equiv|\mbox{\boldmath $A$}|$}. From Eqs.~(\ref{rkinVA}) and
(\ref{phys}) one finds
\begin{equation}
\rho_{\rm kin}=\frac12(\dot A+HA)^2\quad{\rm and}\quad
V_A=\frac12 m^2A^2.
\label{r+V}
\end{equation}

The solution of Eq.~(\ref{eomhom}) is of the form \cite{varkin}
\begin{equation}
A=t^{\frac12(\frac{w-1}{w+1})}\left[c_1J_d(mt)+c_2J_{-d}(mt)\right],
\label{solu}
\end{equation}
where $w$ is the barotropic parameter of the Universe, 
\mbox{$d\equiv\frac{1+3w}{6(1+w)}$}, $c_1,c_2$ are constants of integration
and $J_d$ denotes Bessel function of the the first kind with order $d$.

When the physical vector field is light \mbox{$m\ll H\Leftrightarrow mt\ll 1$}
the above solution can be approximated as \cite{varkin}
\begin{equation}
A=\frac{2}{2w+1}\left(\frac{a}{a_{\rm end}}\right)^{\frac12(3w-1)}
\left(A_{\rm end}+\frac{\dot A_{\rm end}}{H_{\rm end}}\right),
\end{equation}
where the subscript `end' denotes the end of inflation. From the above solution
it can be shown that \cite{varkin}
\begin{equation}
\frac{V_A}{\rho_{\rm kin}}\simeq(mt)^2\ll 1\,.
\end{equation}
Thus, when the vector field is light its energy density is dominated by its
kinetic density. Therefore, \cite{varkin}
\begin{equation}
\rho_A\simeq\rho_{\rm kin}=\frac12
\left(\dot A_{\rm end}+H_{\rm end}A_{\rm end}\right)^2
\left(\frac{a}{a_{\rm end}}\right)^{-4}\Rightarrow\rho_A\propto a^{-4},
\end{equation}
i.e. the light vector field scales as radiation with the Universe expansion.

When the physical vector field is heavy \mbox{$m\gg H\Leftrightarrow mt\gg 1$}
the solution in Eq.~(\ref{solu}) becomes \cite{varkin}
\begin{equation}
A=\sqrt{\frac{2}{\pi}}\,t^{-\frac{1}{1+w}}\left[
c_1\cos\left(mt-\frac{1+2d}{4}\pi\right)+
c_2\cos\left(mt-\frac{1-2d}{4}\pi\right)\right],
\label{heavysolu}
\end{equation}
which shows that the vector field is undergoing rapid quasi-harmonic 
oscillations whose envelope is decreasing as \mbox{$\|A\|\propto a^{-3/2}$}. 
This is easy to understand since, for a heavy vector field, within a Hubble 
time one can ignore the friction term in Eq.~(\ref{eomhom}) and write it as 
\mbox{$\ddot A+m^2 A\simeq 0$}. From Eqs.~(\ref{r+V}) and (\ref{heavysolu})
it is straightforward to find
\begin{equation}
\rho_A=\frac{1}{\pi}\,m^2t^{-\frac{2}{1+w}}\left[
c_1^2+c_2^2+2c_1c_2\cos(d\pi)\right]\Rightarrow\rho_A\propto a^{-3},
\end{equation}
where we used \mbox{$a\propto t^{\frac{2}{3(1+w)}}$} in a spatially flat FRW
Universe. Thus, we see that the density of the heavy oscillating vector field 
scales as pressureless matter with the Universe expansion.

But is it pressureless indeed? From Eqs.~(\ref{rkinp}), (\ref{r+V}) and 
(\ref{heavysolu}) we readily obtain \cite{varkin}
\begin{equation}
\,\!
p_\perp=-\frac{1}{\pi}\,m^2t^{-\frac{2}{1+w}}\left[
c_1^2\sin(2mt-d\pi)+c_2^2\sin(2mt+d\pi)+2c_1c_2\sin(2mt)\right]
\Rightarrow\overline{p_\perp}=0\,,\hspace{-1cm}
\end{equation}
i.e. over a Hubble time (which corresponds to a large number of oscillations)
the average transverse pressure is zero. Since the longitudinal pressure is
$-p_\perp$ it is zero too. This means that the energy-momentum of the 
rapidly oscillating homogeneous vector field is that or pressureless
{\em isotropic} matter (cf. Eq.~(\ref{Tdiag})). Hence, the vector field can 
dominate the Universe without introducing excessive anisotropic stress.
One way of understanding this is that, due to the harmonic oscillations
which send \mbox{$A\rightarrow -A$}, the direction of the vector field is 
rapidly alternated, so that, over a Hubble time, there is 
{\em no net direction}
and the vector field behaves as an approximately isotropic fluid.

\begin{center}
\begin{figure}

\begin{picture}(200,280)
\put(-100,-500){
\leavevmode
\hbox{\epsfxsize=9in
\epsffile{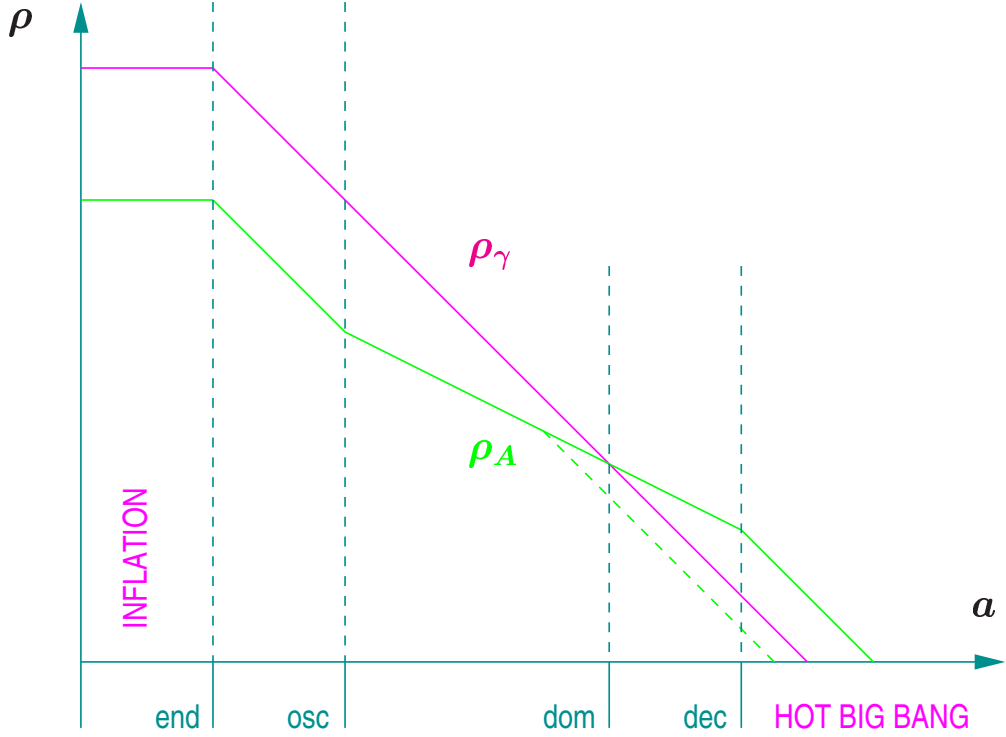%
}}
}

\end{picture}

\caption{\footnotesize
Log-log plot depicting the evolution of the density of the vector curvaton
$\rho_A$ and the background radiation due to inflationary reheating 
$\rho_\gamma$, from the end of inflation and until the onset of the hot big 
bang. During inflation, the vector field has a negligible contribution to the
density of the Universe \mbox{$\rho_A\ll\rho_{\rm inf}$}. At the end of 
inflation (denoted by `end'), the inflationary energy is given to a thermal
bath of radiation \mbox{$\rho_{\rm inf}\rightarrow\rho_\gamma$} (prompt 
reheating is assumed for simplicity). After the end of inflation 
\mbox{$\rho_\gamma\propto a^{-4}$}, which is mimicked by the vector field
density while the vector field remains light, i.e. 
\mbox{$\rho_A\propto a^{-4}$}. Thus, the vector field density parameter 
\mbox{$\Omega_A=\rho_A/\rho$} remains constant with \mbox{$\Omega_A\ll 1$},
where \mbox{$\rho=\rho_\gamma+\rho_A$} is the density of the Universe
(\mbox{$\rho\simeq\rho_\gamma$} during this period). At some later time 
(denoted `osc') the vector field becomes heavy and begins oscillating. From 
then on, it behaves as a pressureless isotropic fluid whose density scales as 
\mbox{$\rho_A\propto a^{-3}$}. Thus, its density parameter grows as 
\mbox{$\Omega_A\propto a$}. This allows the oscillating vector curvaton to 
dominate the Universe at some later moment (denoted by `dom'), when 
\mbox{$\Omega_A\simeq 1$}. Afterwards, the vector curvaton decays into the 
thermal 
bath of the hot big bang (its decay is denoted by `dec'). The slanted dashed 
line corresponds to the possibility when the vector curvaton decays before 
domination, so that \mbox{$\Omega_A^{\rm dec}<1$}. In this case the hot big 
bang begins at the original inflationary reheating.}
\label{fig}
\end{figure}
\end{center}

\subsection{The vector curvaton scenario}

We have seen that the density of a massive Abelian vector field homogenised by 
inflation scales as radiation when the field is light and as matter when it 
becomes heavy. After the end of inflation the energy density is eventually 
transferred into a newly formed thermal bath of relativistic particles. The
density of this thermal bath is dominated by radiation $\rho_\gamma$. The 
homogeneous vector field is initially light so its density also scales as
radiation. Therefore, the density parameter \mbox{$\Omega_A\equiv\rho_A/\rho$} 
of our vector field remains constant, where \mbox{$\rho\approx\rho_\gamma$} is
the density of the Universe. During radiation domination the Hubble parameter
reduces with time as \mbox{$H(t)=1/2t$} so that eventually the vector field
becomes heavy and begins its coherent oscillations. From then on, its density
scales like matter and its density parameter grows \mbox{$\Omega_A\propto a$}.
Thus, the vector field has a chance to dominate (or nearly dominate) the 
Universe before its decay. When it does so it imposes its own curvature 
perturbation onto the Universe, according to the curvaton scenario \cite{curv},
without introducing any anisotropic stress \cite{vecurv}. A schematic 
representation of the vector curvaton scenario is presented in Fig.~\ref{fig}.

The superhorizon perturbations of the vector field satisfy the same equation
of motion as Eq.~(\ref{eomhom}). The reason is that this equation is linear and
that the gradient term is heavily diluted for superhorizon perturbations.%
\footnote{In momentum space the gradient term is $\nabla^2A\rightarrow(k/a)^2A$
where $k/a$ is the physical momentum scale, which is \mbox{$k/a\ll H$} for 
superhorizon perturbations. Note that the mass of the vector field is much
larger than $H$, when it is oscillating and the same is true for its 
perturbations.} Hence, the perturbations follow the same behaviour
as the homogeneous zero-mode (with \mbox{$k=0$}). Thus, when the vector field
becomes heavy they undergo quasi-harmonic oscillations too and their 
anisotropic stress is also eliminated. 

The existence of these perturbations of the vector field implies that the 
density $\rho_A$ is also perturbed and the field's (near) domination of the 
Universe occurs at slightly different times at different locations. This 
results in a difference (perturbation) in the timescale of the Universe 
history, which, according to the $\delta N$ philosophy, results in a 
contribution to the curvature perturbation $\zeta$. Since the density $\rho_A$
is a scalar quantity, this is a {\em scalar} contribution to
$\zeta$ (and not a vector contribution).

Let us now quantify the above. The curvature perturbation in the Universe is,
in principle, the sum of the contribution of the vector field $\zeta_A$ and any
preexisting curvature perturbation, already present in the radiation fluid
$\zeta_\gamma$. Then we can write \cite{curv}
\begin{equation}
\zeta=\zeta_\gamma+\zeta_A=
(1-\hat\Omega_A)\hat\zeta_\gamma+\hat\Omega_A\hat\zeta_A\;,
\end{equation}
where \mbox{$\hat\Omega_A\equiv\frac{3\Omega_A}{4-\Omega_A}\simeq\Omega_A$}
and we have assumed that the vector field has already become heavy. In the 
above, $\hat\zeta_i$ corresponds to the curvature perturbation attributed to 
the $i$-th component of the Universe content, which, on a spatially flat 
slice of spacetime, is given by \cite{curv}
\begin{equation}
\hat\zeta_i\equiv - H\frac{\delta\rho_i}{\dot\rho_i}=
\frac13\frac{\delta\rho_i}{\rho_i+p_i}\,,
\end{equation}
where we used the continuity equation \mbox{$\dot\rho_i+3H(\rho_i+p_i)=0$} for
independent fluids. The above suggest that the contribution of the heavy vector
field to $\zeta$ is
\begin{equation}
\zeta_A=\frac13\hat\Omega_A\frac{\delta\rho_A}{\rho_A}\,.
\end{equation}
At the onset of the oscillations \mbox{$\rho_{\rm kin}\approx V_A$} so that
\mbox{$\rho_A=2V_A=m^2\|A\|^2$}. Thus, to first order we find\footnote{The
zero-mode and the perturbations begin oscillating simultaneously and in phase.}
\begin{equation}
\zeta_A=\frac23\hat\Omega_A\frac{\|A_i\|\,\|\delta A_i\|}{\|A\|^2}=
\frac23\hat\Omega_A\frac{A_i\delta A_i}{A^2}\,.
\end{equation}
From Eq.~(\ref{dN}) we see that the contribution of the vector field to
$\zeta$ to first order is \mbox{$\zeta_A=N_A^i\delta A_i$}. Comparing with the 
above we get
\begin{equation}
N_A^i=\frac23\hat\Omega_A\frac{A_i}{A^2}\Rightarrow
N_A^{ij}=\frac23\hat\Omega_A\frac{\delta_{ij}}{A^2}\,.
\end{equation}
Thus, to second order, the contribution of the vector curvaton to the curvature
perturbation is \cite{stanis}
\begin{equation}
\zeta_A=\frac23\hat\Omega_A\frac{A_i\delta A_i}{A^2}+
\frac13\hat\Omega_A\frac{\delta A_i\delta A_i}{A^2}\,.
\end{equation}
Note, however, that for \mbox{$\delta A_\lambda/A\ll 1$}, the one-loop
correction (last term in the above) is negligible.

Let us now turn our attention to non-Gaussianity. With the above values
of $N_A^i$ and $N_A^{ij}$ and Eqs.~(\ref{fnleql}) and (\ref{fnlsqz}) it
can be shown that \cite{fnlanis}
\begin{equation}
\frac65 f_{\rm NL}^{\rm eql}=\beta^2\calp_+^2\frac{3}{2\hat\Omega_A}
\frac{\left(1+\frac12q^2\right)+\left[p+\frac18\left(p^2-2q^2\right)\right]
\hat A_\perp^2}{(\calp_\phi+\beta\calp_+)^2}
\label{fNLeql}
\end{equation}
and
\begin{equation}
\frac65 f_{\rm NL}^{\rm sqz}=\beta^2\calp_+^2\frac{3}{2\hat\Omega_A}
\frac{1+p\hat A_\perp^2+ipq\,\hat A_\perp\sqrt{1-\hat A_\perp^2}\,
\sin\omega}{(\calp_\phi+\beta\calp_+)^2}\,,
\label{fNLsqz}
\end{equation}
where $p$ and $q$ were defined in Eq.~(\ref{pq}),
$\hat A_\perp$ is the projection of the unit vector 
\mbox{\mbox{\boldmath $\hat A$}$\,=\,$\mbox{\boldmath $A$}$/A$} 
onto the plane of the three momentum vectors
\mbox{\boldmath $k_1$}, \mbox{\boldmath $k_2$} and \mbox{\boldmath $k_3$}
which are used to define the bispectrum (cf. Eq.~(\ref{zB})) and
$\omega$ is the angle between $\hat A_\perp$ and the shorter momentum vector 
in the squeezed configuration.

From Eqs.~(\ref{fNLeql}) and (\ref{fNLsqz}) it is evident that, in the vector 
curvaton scenario with parity invariance, one can write
\begin{equation}
f_{\rm NL}=f_{\rm NL}^{\rm iso}\left(1+{\cal G}\hat A_\perp^2\right)\,
\label{Gdef}
\end{equation}
where $\cal G$ is the anisotropy parameter for non-Gaussianity. In analogy
to $g$ in Eq.~(\ref{gdef}), $\cal G$ quantifies statistical anisotropy in
the bispectrum of $\zeta$. If \mbox{${\cal G}\gg 1$}, then non-Gaussianity is 
predominantly anisotropic, which means that $f_{\rm NL}$ should have a clear 
angular modulation on the sky. If non-Gaussianity is indeed observed and no 
such modulation is found then models which predict \mbox{${\cal G}\gg 1$} will
be ruled out. 

It is important to note above that the directions of statistical anisotropy in 
the spectrum and the bispectrum are correlated, since they are both determined 
by {\boldmath $\hat A$}. This is a smoking gun for the contribution of vector 
fields to $\zeta$.

\section{Particle production of vector fields}\label{pp}

The vector curvaton mechanism can affect (or even generate) the curvature 
perturbation provided the vector field has, somehow, obtained a superhorizon 
spectrum of perturbations during inflation. In order to do this we need to have
a mechanism which breaks the conformal invariance of the vector field. 

A massless Abelian vector field is conformally invariant, which means that it
is not affected by the Universe expansion (it perceives it as a conformal 
transformation to which it is insensitive). Hence, it does not undergo particle
production during inflation. Consequently, its quantum fluctuations do not give
rise to classical perturbations of the field as is done, for example, with 
minimally coupled massless scalar fields. Thus, one expects a light vector 
field to be approximately conformally invariant and the production of its 
perturbations to be suppressed. An explicit breakdown of the vector field 
conformality is, therefore, required. This is model dependent, which suggests
that observations might be able to discern between models and provide insight 
on the underlying theory. 

In this section we discuss two specific models which break the conformality of 
an Abelian vector field and have attracted considerable 
attention to date. Before going into these models however, let us discuss how 
we can use any such mechanism to obtain the spectra of perturbations of the 
vector field components. 

Assume, for the moment, that we are indeed operating under a suitable 
mechanism that breaks the conformality of the vector field. The first step is
to perturb the vector field around the homogeneous value as 
\mbox{$A_\mu\rightarrow A_\mu(t)+\delta A_\mu(\mbox{\boldmath $x$},t)$}.
Then we Fourier transform the perturbations as 
$$\delta\mbox{\boldmath $A$}
(\mbox{\boldmath $k$},t)\equiv
\int\delta\mbox{\boldmath $A$}(\mbox{\boldmath $x$},t)
e^{-i\mbox{\scriptsize\boldmath $k\cdot x$}}d^3x$$
and find the equations of motion of the Fourier components
$\delta$\mbox{\boldmath $A$}$(\mbox{\boldmath $k$},t)$ for the given model.

The next step is to promote the vector field perturbations to
quantum operators by expanding in terms of creation and annihilation operators
\begin{equation}
\delta\mbox{\boldmath $\hat A$}(\mbox{\boldmath $x$},t)
=\int\frac{d^3k}{(2\pi)^3}\sum_\lambda\left[
\mbox{\boldmath $\hat e$}^\lambda\hat a_\lambda(\mbox{\boldmath $k$})
\delta\cala(k,t)e^{i\mbox{\scriptsize\boldmath $k\cdot x$}}+
\mbox{\boldmath $\hat e$}^{\lambda*}\hat a_\lambda^\dag(\mbox{\boldmath $k$})
\delta\cala^*(k,t)e^{-i\mbox{\scriptsize\boldmath $k\cdot x$}}\right].
\end{equation}
The mode functions $\delta\cala(k,t)$ and the Fourier components of the 
perturbations of the vector field satisfy the same equations of motion because 
the latter are linear. Thus, we need to solve these equations and find the
the mode functions. To do this we need to employ the following
boundary conditions, which are simply due to the fact that the perturbations
begin as quantum fluctuations well within the horizon 
(\mbox{$k/aH\rightarrow\infty$})
\begin{equation}
{\delta\cala_{L,R}}_{_{_{_{\hspace{-1cm}k/aH\rightarrow\infty}}}}
=\frac{e^{ik/aH}}{\sqrt{2k}}\quad{\rm and}\quad
{\delta\cala_\|}_{_{_{_{\hspace{-.7cm}k/aH\rightarrow\infty}}}}
\!\!\!\!=\gamma\;\frac{e^{ik/aH}}{\sqrt{2k}},
\label{BD}
\end{equation}
where \mbox{$\gamma\equiv\frac{E}{m}=\sqrt{(\frac{k}{am})^2+1}$} is the Lorentz
boost factor which takes us from the frame with 
\mbox{\mbox{\boldmath $k$}$\,=0$}
(where all components of the vector field perturbation are equivalent) to
the one of momentum \mbox{\boldmath $k$}. Apart from $\gamma$, we see that the
vacuum boundary conditions are identical to the Bunch-Davis vacuum also 
employed for the particle production of scalar fields.

Once we solve the equations of motion and find the mode functions we can obtain
the power spectra of the superhorizon (\mbox{$k/aH\rightarrow 0$}) 
perturbations using
\begin{equation}
\calp_\lambda=\frac{k^3}{2\pi^2}
{|\delta\cala_\lambda|^2}_{_{_{_{\hspace{-1cm}k/aH\rightarrow 0}}}}\;.
\label{Pl}
\end{equation}
The typical value of the vector field perturbation is 
\mbox{$\delta A_\lambda\sim\sqrt{\calp_\lambda}$}. Now, let us employ this 
method on two concrete models for the generation of a perturbation spectrum for
the vector field during inflation.

\subsection{Non-minimal coupling to gravity}

This mechanism was first considered in Ref.~\cite{TW} for the generation of 
a primordial magnetic field of superhorizon coherence. It was employed as
a vector curvaton in Ref.~\cite {nonmin} and also in Ref.~\cite{stanis}.

Consider a massive Abelian vector field with a non-minimal coupling to gravity
as follows
\begin{equation}
{\cal L}=-\frac14 F_{\mu\nu}F^{\mu\nu}+\frac12(m^2+\alpha R)W_\mu W^\mu,
\label{LR}
\end{equation}
where $R$ is the scalar curvature and $\alpha$ is a constant. The non-minimal
coupling corresponds to a contribution to the effective mass of the field such
that
\begin{equation}
m_{\rm eff}^2\equiv m^2+\alpha R\,.
\end{equation}

The equations of motion have been found and solved for the above theory during 
de Sitter inflation, 
providing the following exact solutions for the mode functions 
$\delta\cala_\lambda$ of the perturbations. For the transverse components the
solution is \cite{nonmin}
\begin{equation}
\delta\cala_{L,R}=a^{-3/2}\sqrt{\frac{\pi}{H}}
\frac{e^{i\frac{\pi}{2}(\nu-\frac12)}}{1-e^{i2\pi\nu}}\left[
J_\nu\left(\frac{k}{aH}\right)-e^{i\pi\nu}
J_{-\nu}\left(\frac{k}{aH}\right)\right],
\end{equation}
where
\begin{equation}
\nu\equiv\sqrt{\frac14-\left(\frac{m_{\rm eff}}{H}\right)^2}.
\end{equation}
The above solution produces a scale invariant spectrum if \mbox{$\nu=3/2$}.
This can be achieved if \mbox{$m\ll H$} and \mbox{$\alpha\approx\frac16$},
because during de Sitter inflation \mbox{$R=-12H^2$}.%
\footnote{This theory is in effect a modified gravity theory but it can be
shown that the Friedman Equation is not affected if \mbox{$\alpha=\frac16$}
and also $RW^2$ is negligible compared to the Einstein-Hilbert action for
\mbox{$W_\mu W^\mu\ll m_P^2$}.} With this choice
for $m$ and $\alpha$ the solution for the longitudinal mode function is 
\cite{stanis}
\begin{equation}
\delta\cala_\|=\frac{1}{\sqrt 2}\left[\left(\frac{k}{aH}\right)-2
\left(\frac{aH}{k}\right)+2i\right]\frac{e^{ik/aH}}{\sqrt{2k}}\,.
\end{equation}

The transverse solutions are the same because the theory is parity invariant.
However, it is clear that there is striking difference between the transverse 
and the longitudinal solutions. Using Eq.~(\ref{Pl}), we can now find the
power spectra for the components of the perturbation. We obtain
\begin{equation}
\calp_+=\left(\frac{H}{2\pi}\right)^2,\quad\calp_-=0\quad{\rm and}\quad
\calp_\|=2\left(\frac{H}{2\pi}\right)^2,
\end{equation}
i.e. \mbox{$p=1$} and \mbox{$q=0$} as expected (cf. Eq.~(\ref{pq})). 
Thus we find that particle production is anisotropic at a level of 
100\%. This means that the vector field contribution to $\zeta$ should be
subdominant, for otherwise it would violate the observational constrains which 
do not allow statistical anisotropy above 30\%. Hence, we have to assume that
$\zeta$ is primarily due to some other source, presumably a light scalar field,
and the contribution of the vector field is significant only at the level of 
generating significant statistical anisotropy. 

From Eq.~(\ref{g}), we obtain for the anisotropy parameter
\begin{equation}
g=\frac{\beta}{1+\beta}\approx\beta\ll 1\,
\end{equation}
where we considered that \mbox{$\calp_\phi=(H/2\pi)^2$} and also that it is the
scalar field which primarily modulates $N$ so that \mbox{$\beta\ll 1$} in
Eq.~(\ref{beta}).

Using that \mbox{$p=1$} and \mbox{$q=0$} in Eqs~(\ref{fNLeql}) and 
(\ref{fNLsqz}), we obtain
\begin{equation}
\frac65 f_{\rm NL}^{\rm eql}=2\frac{\beta^2}{\Omega_A}
\left(1+\frac98\hat A_\perp^2\right)\quad{\rm and}\quad
\frac65 f_{\rm NL}^{\rm sqz}=2\frac{\beta^2}{\Omega_A}
\left(1+\hat A_\perp^2\right).
\end{equation}
Thus, we see that \mbox{${\cal G}\sim 1$}. This is because, in this theory, 
there is only one mass-scale involved, that is $H$. Dimensionless quantities, 
therefore, such as $\cal G$ or $p$ are expected to be of order unity. In 
Ref.~\cite{fnlanis} it was shown that, whenever it is so, there is a clear 
prediction for the maximum non-Gaussianity, which provides a direct link
with statistical anisotropy in the spectrum:
\begin{equation}
f_{\rm NL}^{\rm max}\sim 10^3\left(\frac{g}{0.1}\right)^{3/2}.
\end{equation}
From the above, it is evident that, through the vector curvaton mechanism, 
significant non-Gaussianity can be produced.

This theory was criticised in that it may suffer from instabilities such as 
ghosts \cite{peloso}. However, it is not clear whether this is indeed so.
In Ref.~\cite{peloso}, it was shown that the modes of the longitudinal 
perturbations are ghosts but only when subhorizon. Given that these modes
are subhorizon only for a limited time and also in view of the fact that the
energy density of inflation is much larger than $\rho_A$, one may wonder 
whether these ghosts manage to destabilise the vacuum. A discussion on this 
issue can be found in Ref.~\cite{M+D}.

\subsection{Varying kinetic function and mass}

Consider a massive Abelian vector boson field, with Lagrangian density
\begin{equation}
{\cal L}=-\frac14 fF_{\mu\nu}F^{\mu\nu}+\frac12 m^2W_\mu W^\mu,
\label{L}
\end{equation}
where \mbox{$f=f(t)$} is the kinetic function, which is approaching unity
by the end of inflation so that, afterwards, the vector field is canonically 
normalised. The above theory does not suffer from instabilities (ghost free)
\cite{insta,insta+}, which motivates the use of the Maxwell-type kinetic term
even if the field is not a gauge boson. Note here that an Abelian massive 
vector field is renormalisable even if it is not a gauge boson \cite{tikto}.

The spatial components of the physical canonically normalised vector field in 
this case are \cite{varkin}
\begin{equation}
\mbox{\boldmath $A$}=\sqrt f\,\mbox{\boldmath $W$}/a\,.
\label{phys+}
\end{equation}
The mass of the physical, canonically normalised vector field is
\begin{equation}
M\equiv\frac{m}{\sqrt f}\,.
\label{M}
\end{equation}

The massless version of this theory has been extensively considered, firstly 
for inflationary particle production for the generation of a primordial
magnetic field \cite{gaugekin} and more recently for the mild anisotropisation
of inflation, which gives rise to statistical anisotropy in the curvature
perturbation through anisotropic particle production of the inflaton 
\cite{anisinf}. Here we 
investigate if this vector field can play the role of the vector curvaton.
This was first studied in Ref.~\cite{sugravec}.

The equations of motion for the mode functions for this model were obtained in
Ref.~\cite{sugravec}. They read
\begin{equation}
\left[\partial_t^2+\left(H+\frac{\dot f}{f}\right)\partial_t
+\left(\frac{k}{a}\right)^2+\frac{m^2}{f}\right]
\delta {\cal W}_{L,R}=0
\end{equation}
and
\begin{equation}
\left[\partial_t^2+\left(H+\frac{\dot f}{f}\right)\partial_t
+\left(2H+2\frac{\dot m}{m}-\frac{\dot f}{f}\right)
\frac{\left(\frac{k}{a}\right)^2\partial_t}{\left(\frac{k}{a}\right)^2+
\frac{m^2}{f}}+\left(\frac{k}{a}\right)^2+\frac{m^2}{f}\right]
\delta {\cal W}_\|=0\,,
\end{equation}
where \mbox{$\delta{\cal W}_\lambda=a\,\delta\cala_\lambda/\sqrt f$}, cf. 
Eq.~(\ref{phys+}). Using the above, particle production has been studied in
Ref.~\cite{varkin}. It was found that, the transverse components obtain
a scale invariant superhorizon spectrum of perturbations when
\begin{equation}
f\propto a^{-1\pm 3}\quad{\rm and}\quad
M_*\ll H_*
\label{trans}
\end{equation}
where the subscript `*' denotes the time of horizon exit. To obtain a 
scale-invariant spectrum for the longitudinal component one needs an additional
condition on the time-dependence of $m(t)$, which reads \cite{varkin}
\begin{equation}
m\propto a\,.
\end{equation}

If the vector field is a gauge boson then $f$ is the gauge kinetic function 
which is related with the gauge coupling as \mbox{$f\sim 1/e^2$}. This means 
that only the case when \mbox{$f\propto a^{-4}$} is possible since, only then 
does the gauge field remain weakly coupled during inflation. The gauge kinetic
function is one of the three fundamental functions\ which define a supergravity
 theory.%
\footnote{The other two are the K\"ahler potential and the superpotential.}
In supergravity, the gauge kinetic function is a holomorphic function of the 
scalar fields of the theory. Now, during inflation, supergravity corrections
are expected to give masses $\sim H$ to the scalar fields \cite{randall}.
This means that, during inflation, these scalar fields are fast-rolling down 
the slopes of the scalar potential, which would cause significant variation to
the gauge kinetic function. Indeed, it is easy to show that 
\mbox{$\dot f/f\sim H$} is natural to expect. Here we should note that,
if $f$ is modulated by the inflaton field then, under fairly general 
conditions, \mbox{$f\propto a^{-4}$} is an attractor solution during 
inflation,
which arises due to the backreaction of the vector field onto the roll of the 
inflaton down the inflationary scalar potential \cite{attract}. Thus, even 
though, originally, the inflaton may be also fast-rolling, the backreaction
slows down its roll and allows slow-roll inflation to occur even with a 
relatively steep scalar potential. Indeed, it was shown in Ref.~\cite{attract}
that this is a neat way to overcome the infamous $\eta$-problem of inflation,
while simultaneously obtaining \mbox{$f\propto a^{-4}$} as an attractor 
solution.

Now let us discuss the behaviour of the mode functions $\delta\cala_\lambda$
in more detail. Firstly, we define
\begin{equation}
x\equiv\frac{k}{aH}\quad{\rm and}\quad z\equiv\frac{M}{3H}\,.
\label{xz}
\end{equation}
From the above and Eq.~(\ref{M}) it is evident that, if \mbox{$f\propto a^2$}
then \mbox{$z=\,$constant}, while if \mbox{$f\propto a^{-4}$} then
\mbox{$z\propto a^3$}. Note that \mbox{$x\propto a^{-1}$} always, while
Eq.~(\ref{trans}) requires that \mbox{$z_*\ll 1$}.

It turns out that there are three possible stages for the mode evolution 
\cite{varkin}.
When \mbox{$x>1\gg z$}, then the mode is still subhorizon and it is oscillating
so that it can be matched to the boundary conditions in Eq.~(\ref{BD}). As time
passes $x$ decreases and the mode becomes superhorizon. When \mbox{$x,z\ll 1$}
then the mode is found to undergo power-law evolution. The third possible stage
has to do with the case \mbox{$f\propto a^{-4}$} only, when $z$ is growing in 
time. In this case we could finally reach the time when \mbox{$z\gsim 1\gg x$},
when the superhorizon mode begins oscillating again. Since we are interested 
in superhorizon scales (they are the ones which can affect the curvature 
perturbation as observed in the CMB) we will consider the modes which are
caught by the end of inflation either when \mbox{$x,z\ll 1$}, or when 
\mbox{$z\gsim 1\gg x$}, which is possible only with  \mbox{$f\propto a^{-4}$}.

\subsubsection{Power-law regime}

This is the case when \mbox{$M<3H\Leftrightarrow z<1$} when inflation ends. 
The mode functions for the superhorizon modes are found to be \cite{varkin}
\begin{equation}
\delta\cala_{L,R}=\frac{i}{\sqrt{2k}}\left(\frac{H}{k}\right)
\quad{\rm and}\quad
\delta\cala_\|=-\frac{1}{\sqrt{2k}}\left(\frac{H}{k}\right)\frac{1}{z}\,.
\end{equation}
Using this, the power spectra for the superhorizon perturbations of the vector 
field are 
\begin{equation}
\calp_+=\calp_{L,R}=\left(\frac{H}{2\pi}\right)^2
\quad{\rm and}\quad
\calp_\|=\left(\frac{H}{2\pi}\right)^2\left(\frac{H}{3M}\right)^2,
\end{equation}
where we used that the theory is parity invariant. Because \mbox{$M<3H$} in
this regime we find that \mbox{$\calp_+\ll\calp_\|$}, which means that particle
production is strongly anisotropic. Therefore, if inflation ends in this 
power-law regime then the contribution of the vector curvaton to the curvature
perturbation has to be subdominant (otherwise it would generate excessive 
statistical anisotropy), so that \mbox{$\beta\ll 1$}, which means that $N$ is
primarily modulated by a scalar field and not by our vector curvaton. In this
case, therefore, the vector curvaton can only generate some statistical 
anisotropy in $\zeta$. For the spectrum, the anisotropy parameter is 
(cf. Eq.~(\ref{g}))
\begin{equation}
g=\beta\frac{\calp_\|}{\calp_+}=\beta/z^2,
\end{equation}
where \mbox{$\calp_\phi=\calp_+$}.
Similarly, for the non-linearity parameter we obtain
\begin{equation}
\frac65 f_{\rm NL}=\beta^2\frac{3}{2\hat\Omega_A}
\left[1+\left(p+\frac18\kappa p^2\right)\hat A_\perp^2\right],
\end{equation}
where \mbox{$\kappa=1$} \{\mbox{$\kappa=0$}\} for the equilateral \{squeezed\}
configurations and we used Eqs.~(\ref{fNLeql}) and (\ref{fNLsqz}) and also
that \mbox{$\beta\ll 1$}, \mbox{$q=0$} and \mbox{$p=1/z^2$}
(cf. Eq.~(\ref{pq})). Since the vector curvaton must have a subdominant 
contribution to $\zeta$ we have \mbox{$\Omega_A\ll 1$}, which gives
\mbox{$\hat\Omega_A\rightarrow\frac34\Omega_A$}. Using this and the above, we
find that the isotropic part of non-Gaussianity has
\begin{equation}
f_{\rm NL}^{\rm iso}=\frac53\frac{\beta^2}{\Omega_A}
=\frac53\frac{g^2z^4}{\Omega_A}\,,
\end{equation}
while the anisotropy parameter for non-Gaussianity is
\begin{equation}
{\cal G}=p+\frac18\kappa p^2\gg 1\,,
\end{equation}
i.e. non-Gaussianity is predominantly anisotropic. Thus, if non-Gaussianity is 
indeed observed and it does not feature a strong angular modulation this regime
of this model will be ruled out.\footnote{Note here that \mbox{${\cal G}\gg 1$}
because we have two scales into the model, $H$ and $M$, in contrast to the
non-minimally coupled model of the previous section, where we only had one 
scale $H$ so that \mbox{${\cal G}\sim 1$}.}

If \mbox{$f\propto a^2$} then \mbox{$M=\,$constant} and, because of 
Eq.~(\ref{trans}), \mbox{$M=M_*\ll H$}. Thus, in this case 
\mbox{$\calp_+,\calp_\|=\,$constant} and the above are the only possibility.
If, however, \mbox{$f\propto a^{-4}$} then \mbox{$M\propto a^3$} and
$\calp_\|$ is decreasing in time. Yet, scale invariance is maintained because
the amplitude of the modes when they exit the horizon is reduced in time in
accordance to the reduction of the value of the spectrum. As a result, the 
piling of modes does not spoil the flatness of the superhorizon spectrum
\cite{varkin}. This case offers the possibility that $M$
will reach and surpass $H$ before the end of inflation. If this happens, then
we are no more in the power-law regime.

\subsubsection{Oscillatory regime}

This regime corresponds to the possibility that 
\mbox{$M\gg H\Leftrightarrow z\gg 1$} when inflation 
ends. Because of Eq.~(\ref{trans}), when the cosmological scales exit the 
horizon we have \mbox{$M_*\ll H$}. Thus, this regime can be realised only if
$M$ is growing during inflation, which is possible only if 
\mbox{$f\propto a^{-4}$}. Note that, when \mbox{$M\gsim H$}, 
particle production
stops and the perturbation of the field on scales that leave the horizon 
becomes essentially zero (exponentially suppressed). However, the superhorizon 
scales which left the horizon when the vector 
field was still light, retain their perturbations, which evolve as follows.

In this case, the mode functions are given by \cite{varkin}
\begin{eqnarray}
2\sqrt{\frac{H}{\pi}}\left(\frac{k}{H}\right)^{3/2}\delta\cala_{L,R} & = &
\frac{i}{\sqrt z}J_{1/2}(z)+\frac13 x^3\sqrt zJ_{-1/2}(z)\\
 & & \nonumber\\
2\sqrt{\frac{H}{\pi}}\left(\frac{k}{H}\right)^{3/2}\delta\cala_\| & = &
\frac13 x^3\sqrt zJ_{1/2}(z)-\frac{1}{\sqrt z}J_{-1/2}(z)\,.
\end{eqnarray}
Considering superhorizon modes (i.e. \mbox{$x\rightarrow 0$}) and also that
\mbox{$z\gg 1$} in this regime, the above simplify to
\begin{equation}
\delta\cala_{L,R}=\frac{i}{\sqrt{2H}}\left(\frac{H}{k}\right)^{3/2}
\frac{\sin z}{z}
\quad{\rm and}\quad
\delta\cala_\|=-\frac{1}{\sqrt{2H}}\left(\frac{H}{k}\right)^{3/2}
\frac{\cos z}{z}\,.
\end{equation}
Thus, we see that the mode functions are undergoing rapid coherent oscillations
since \mbox{$z\gg 1$}. Using these, we find that the average of the power 
spectra is
\begin{equation}
\overline{\calp_+}=\overline{\calp_\|}=
\frac12\left(\frac{H}{2\pi}\right)^2\left(\frac{H}{3M}\right)^2.
\end{equation}
Thus, we see that we have isotropic particle production, which implies 
\mbox{$p=q=0$} (cf. Eq.~(\ref{pq})). In this case we do 
not need the input of a scalar field to generate the curvature perturbation as
$\zeta$ can indeed be fully produced by the vector field alone. To our 
knowledge this is the only model that manages to produce $\zeta$ without the 
direct involvement of a fundamental scalar field. Because we need no scalar 
field, \mbox{$\beta\gg 1$}. Then Eqs.~(\ref{fNLeql}) and (\ref{fNLsqz}) both
reduce to
\begin{equation}
f_{\rm NL}=\frac{5}{4\hat\Omega_A}\,,
\end{equation}
which is identical to the case of a scalar curvaton \cite{curv}.

\subsubsection{Borderline regime}

It is interesting to briefly consider the so-called ``borderline'' regime,
when \mbox{$M\sim 3H\Leftrightarrow z\sim 1$} at the end of inflation.
Again, this is realisable only if $M$ is growing during inflation, i.e. when
\mbox{$f\propto a^{-4}$}.

In this case one has
\begin{equation}
g=p=\frac{\calp_\|}{\calp_+}-1\equiv\frac{\delta\calp}{\calp_+}\,,
\end{equation}
where \mbox{$\delta\calp\equiv\calp_\|-\calp_+$}. Thus, we see that, to satisfy
observational bounds \mbox{$\calp_\|\approx\calp_+$} at least within 30\%.
If this is the case then it can be shown that \cite{varkin}
\begin{equation}
f_{\rm NL}=\frac{5}{4\hat\Omega_A}\left(1+g\hat A_\perp^2\right).
\end{equation}
Therefore, statistical anisotropy in the spectrum and the non-Gaussianity have
the same magnitude, i.e. \mbox{${\cal G}=g$}. This is an interesting 
characteristic signature for this scenario.

\subsubsection{The evolution of the zero mode}

When \mbox{$f\propto a^{-1\pm 3}$}, it is straightforward to show that the 
equation of motion for the homogeneous physical vector field is
\begin{equation}
\mbox{\boldmath $\ddot A$}+3H\mbox{\boldmath $\dot A$}+M^2\mbox{\boldmath $A$}
=0\,,
\label{eomA}
\end{equation}
which looks identical to the Klein-Gordon equation for a minimally coupled 
massive scalar field. 

If \mbox{$f\propto a^2$} then \mbox{$M=\,$constant$\,\ll 1$} and we have
\mbox{\mbox{\boldmath $A$}$\,\simeq\,$constant}. This means that
\begin{equation}
\rho_A\simeq V_A\sim M_0^2A_0^2={\rm constant}\,,
\end{equation}
where $M_0$ is the initial
value of $M$ (\mbox{$M=M_0$}, since $M$ is constant) and $A_0$ is the initial 
value of \mbox{$A=|\mbox{\boldmath $A$}|$}.

Now, if \mbox{$f\propto a^{-4}$} then \mbox{$M\propto a^3$} and the solution
to Eq.~(\ref{eomA}) is \cite{varkin}
\begin{equation}
A=A_0\left(\frac{a}{a_0}\right)^{-3}\sqrt 2\cos\left(z\pm\frac{\pi}{4}\right),
\end{equation}
which means that the typical value of the vector field is 
\mbox{$A\propto a^{-3}$}. Using this, Eq.~(\ref{r+V}) gives
\begin{equation}
\rho_{\rm kin}=\left[A_0M_0\sin\left(z\pm\frac{\pi}{4}\right)\right]^2
\quad{\rm and}\quad
V_A=\left[A_0M_0\cos\left(z\pm\frac{\pi}{4}\right)\right]^2,
\end{equation}
which results in
\begin{equation}
\rho_A=\rho_{\rm kin}+V_A=M_0^2A_0^2={\rm constant}\,.
\end{equation}
Thus, we see that if \mbox{$f\propto a^{-1\pm 3}$} then 
\mbox{$\rho_A=\,$constant} during inflation.

\subsubsection{Curvaton physics}

Let us briefly look into how the model parameters are constrained by the 
requirement that the vector field performs as a successful curvaton. As shown
in Refs.~\cite{varkin,insta} such considerations impose the following
constraint on the model parameters
\begin{equation}
\frac{H_*}{m_P}\sim\hat\zeta_A\sqrt{\Omega_A^{\rm dec}}
\left(\frac{\max\{\Gamma_A;H_{\rm dom}\}}{\min\{m_A;H_*\}}\right)^{1/4},
\label{cons}
\end{equation}
where `dec' denotes the moment of the vector curvaton decay, `dom' denotes
the moment when the vector curvaton dominates the Universe (if it does not 
decay earlier than that), $\Gamma_A$ is the decay rate of the vector curvaton
and $m_A$ is the final value of its mass at the end of inflation.

If the vector field is still light at the end of inflation, i.e. if
\mbox{$m_A\ll H_*$}, we are in the anisotropic regime. Then we obtain
\cite{varkin}
\begin{equation}
\zeta\sim\frac{\Omega_A^{\rm dec}\hat\zeta_A}{\sqrt g}\,.
\end{equation}
Using this and also the requirement that the curvaton decays before 
Big Bang Nucleosynthesis (BBN) (i.e. \mbox{$\Gamma_A>T_{\rm BBN}^2/m_P$}
where \mbox{$T_{\rm BBN}\sim 1\,$MeV} is the temperature at BBN) the constraint
in Eq.~(\ref{cons}) gives
\begin{equation}
H_*>\sqrt g\times 10^7\,{\rm GeV}
\quad{\rm and}\quad
10\,{\rm Tev}\lsim m_A\ll H_*\;.
\label{range1}
\end{equation}

If the vector field becomes heavy by the end of inflation, i.e. if
\mbox{$m_A\gsim H_*$}, then we are in the (almost) isotropic regime. 
In this case the vector curvaton alone can be responsible for $\zeta$
as we have discussed. Thus, we have \cite{varkin}
\begin{equation}
\zeta=\zeta_A\sim\Omega_A^{\rm dec}\hat\zeta_A
\Rightarrow\Omega_A^{\rm dec}H_*\gsim\zeta^2m_P\;.
\end{equation}
This results in the following bound
\begin{equation}
H_*\gsim 10^9\,{\rm GeV}\,.
\end{equation}
The above corresponds to a relatively high scale of inflation which, in 
supergravity models, may result in gravitino overproduction. However, the 
latter can be avoided through the entropy release by the vector curvaton decay 
which can dilute the gravitinos.

Taking into account that the vector curvaton can decay at least through 
gravitational couplings, we find \mbox{$\Gamma_A\geq m_A^3/m_P^2$}. Using this
it can be shown that the oscillations of the massive vector field cannot
commence earlier than \mbox{$N_{\rm osc}^{\rm max}\simeq 4.4$} e-folds of
inflation. As a result, the range of $m_A$ for which this regime can be 
realised is found to be \cite{varkin}
\begin{equation}
1<\frac{m_A}{H_*}<10^6.
\label{range2}
\end{equation}
From Eqs.~(\ref{range1}) and (\ref{range2}) we see that there is ample 
parameter space for both the possibilities (light or heavy vector curvaton) to 
be realised. 

\section{Conclusions}

The high precision cosmological observations enable cosmologists to 
investigate beyond the ``vanilla'' predictions of inflation and thereby
discriminate between inflation models. A new such observable is statistical 
anisotropy, which amounts to direction dependent patterns in the CMB (and
possibly large scale structure too). Statistical anisotropy is within the reach
of the forthcoming observations of the Planck satellite mission, which is 
expected to release its first cosmological data in the beginning of 2013. 
Currently, the latest CMB observations allow statistical anisotropy in the 
spectrum as much as 30\% (\mbox{$g\lsim 0.3$}). Planck will reduce this bound 
down to 2\% if statistical anisotropy is not observed. It should be noted here 
that, even if the spectrum is weakly statistically anisotropic, the bispectrum 
can be predominantly anisotropic with \mbox{${\cal G}\gg 1$}. This means that, 
if non-Gaussianity is indeed observed without a strong angular modulation of
$f_{\rm NL}$ on the microwave sky, then all models which predict  
\mbox{${\cal G}\gg 1$} will be ruled out. 

Vector boson fields are natural candidates for the generation of statistical 
anisotropy in the curvature perturbation, because they are expected to pick a 
preferred direction when homogenised by inflation. The vector curvaton 
paradigm offers a simple, generic and effective mechanism for the direct 
contribution of vector boson fields to the curvature perturbation. The 
mechanism assumes a Proca vector field, whose zero-mode begins oscillating when
the field becomes heavy, after the end of inflation. As shown, the oscillating 
zero-mode behaves as a pressureless isotropic fluid and can (nearly) dominate 
the Universe without generating an excessive anisotropic stress. When doing so 
it imposes its own contribution to the curvature perturbation, in accordance to
the curvaton mechanism. We should point out here that this perturbation is 
scalar in nature, because it is due to the perturbed value of the density of 
the vector field, which is a scalar quantity. A considerable advantage of the
vector curvaton mechanism is that it does not rely on an interaction of any 
kind between the vector field and the inflaton sectors. This allows the vector 
field to correspond to physics at a much lower energy scale (e.g. TeV physics)
than the scale of inflation, which may connect with observations in collider 
experiments such as the LHC.

The particle production process, through which the vector curvaton obtains a
superhorizon spectrum of perturbations, is in general anisotropic. This is
because the vector boson field has more than one degrees of freedom (three if 
massive), for which the efficiency of the particle production is in general 
different. Thus, the curvature perturbation contributed by a vector curvaton
is, in general, statistically anisotropic. If particle production is indeed
isotropic (at least at the level allowed by the observations) the vector 
curvaton mechanism can generate the curvature perturbation in the Universe from
vector fields alone without directly involving any fundamental scalar fields.
If, however, statistical anisotropy is indeed observed, then we {\em have} to 
go beyond scalar fields to explain the observations. This means that, the 
observation of statistical anisotropy in the CMB, may probe the gauge field 
content of theories beyond the standard model. 

There are some related issues which we did not go into in this paper. For 
example, studies of the trispectrum in vector field scenarios \cite{Ytri},
or of one-loop contributions \cite{Y1loop}. Note also, the possibility that the
vector curvaton is non-Abelian is investigated in Ref.~\cite{nonAbelian}, while
the contribution to the curvature perturbation by P-forms can be found in 
Ref.~\cite{Pforms}.

The interest now is in finding realistic candidates in theories beyond the 
standard model, which can play the role of the vector curvaton. Examples can 
include the supermassive gauge bosons of grand unified theories 
\cite{WDL}\footnote{These can be associated with the simultaneous generation of
a primordial magnetic field \cite{pmf}.} 
or the vector fluxes on probe branes in the context of DBI-inflation 
\cite{DBIvc}. Another promising idea is investigating the possibility that the 
vector bosons associated with gauged axions can play the role of the vector 
curvaton. In this case, the generation of parity-violating statistical 
anisotropy is possible \cite{anomaly}.

\section*{Acknowledgements}

I would like to thank my collaborators Mindaugas Kar\v{c}iauskas, David H. 
Lyth, Yeinzon Rodriguez-Garcia and Jacques M. Wagstaff. I am grateful to the 
University of Crete for the hospitality.



\end{document}